\documentclass[fleqn,twoside]{article}
\usepackage{espcrc2}

\usepackage{graphicx}
\usepackage[figuresright]{rotating}

\newcommand{\cf}{{\it cf.}}

\newcommand{\eq}{Eq.}

\newcommand{\fig}{Fig.}

\newcommand{\Ref}{Ref.}

\newcommand{\stheta}{\sin^22\theta_{13}}
\newcommand{\deltacp}{\delta_\mathrm{CP}}
\newcommand{\ldm}{\Delta m_{31}^2}

\newcommand{\equ}[1]{\eq~(\ref{equ:#1})}
\newcommand{\figu}[1]{\fig~\ref{fig:#1}}

\newcommand{\bi}{\begin{itemize}}
\newcommand{\ei}{\end{itemize}}

\newcommand{\AmS}{{\protect\the\textfont2
  A\kern-.1667em\lower.5ex\hbox{M}\kern-.125emS}}

\title{Geographical issues and physics applications of ``very'' long NF baselines}

\author{Walter Winter\address{School of Natural Sciences, Institute for Advanced Study,
        Princeton, NJ 08540}%
        \thanks{Work supported by the W.~M.~Keck Foundation and NSF grant PHY-0070928.}}

\begin{document}

\begin{abstract}
We discuss several potential applications of ``very'' long neutrino factory (NF)
baselines, as well as potential detector locations for these applications.
\vspace{-1pc}
\end{abstract}

\maketitle

Neutrino factories~\cite{Geer:1998iz} are usually discussed in combination with baselines
up to about $3 \, 000 \, \mathrm{km}$.  However, one could think about potential applications
of a second, much longer baseline, which could be operated simultaneously or subsequently.
In the following, we refer to ``very'' long baselines as baselines much longer than $3 \, 000 \, \mathrm{km}$.

For long-baseline beam experiments, the electron or muon neutrino appearance probability $P_{\mathrm{app}}$ ($P_{e \mu}$, $P_{\mu e}$, $P_{\bar{e} \bar{\mu}}$, or $P_{\bar{\mu} \bar{e}}$) is very sensitive to matter effects. It can be expanded in the small hierarchy parameter $\alpha \equiv \Delta m_{21}^2/\Delta m_{31}^2$ and the small $\sin 2 \theta_{13}$ up to the second order as (see \Ref~\cite{Akhmedov:2004ny} and references therein):
\begin{eqnarray}
P_{\mathrm{app}} & \simeq & \Delta^2 \cdot \left( \sin^2 2\theta_{13} \cdot  \sin^2 \theta_{23} \cdot f_1^2 \right.
\nonumber \\
&\pm&   \alpha  \cdot \sin 2\theta_{13} \cdot \sin \delta_{\mathrm{CP}} \cdot
 \sin \Delta \cdot \xi \cdot f_1 \cdot f_2
\nonumber  \\
&+&   \alpha  \cdot \sin 2\theta_{13} \cdot   \cos \delta_{\mathrm{CP}} \cdot \cos \Delta \cdot \xi \cdot f_1 \cdot f_2
 \nonumber  \\
&+&  \left. \alpha^2 \cdot \cos^2 \theta_{23}  \cdot \sin^2 2\theta_{12} \cdot f_2^2 \right) \, .
\label{equ:PROBMATTER}
\end{eqnarray}
Here $\Delta \equiv \Delta m_{31}^2 L/(4 E)$, $\xi = \sin 2\theta_{12} \cdot \sin 2\theta_{23}$,
\begin{equation}
f_1  = \frac{\sin[(1-\hat{A}){\Delta}]}{[(1-\hat{A}) \Delta ]} \, , \quad
f_2  = \frac{\sin(\hat{A}{\Delta})}{(\hat{A} \Delta)} \, ,
\end{equation}
 and $\hat{A} \equiv \pm (2 \sqrt{2} G_F n_e E)/\Delta m_{31}^2$.
Note that the matter effect in \equ{PROBMATTER} enters via the matter potential $\hat{A}$, where the
equation reduces to the vacuum case for $\hat{A} \rightarrow 0$. In addition, the combination $\hat{A} \Delta = \sqrt{2}/2 \, G_F n_e L$ does not depend on energy or oscillation parameters.

From \equ{PROBMATTER}, we can read off a special structure in terms of the factors $f_1$ and
$f_2$: The first term is proportional to $f_1^2$, the second and third to $f_1 \cdot f_2$, and the fourth to $f_2^2$. In addition, we have pulled out a factor of $\Delta^2 \propto L^2$ from the equation, which means that the $1/L^2$ geometrical drop of the flux is compensated by this factor, and that $f_1$ and $f_2$ determine the individual weights of the four terms as function of
baseline, energy, and $\ldm$. However, note that the relative weight between the second and third CP terms is in addition given by the vacuum oscillation phase $\Delta$.
\begin{figure}[t]
\begin{center}
\includegraphics[width=\columnwidth]{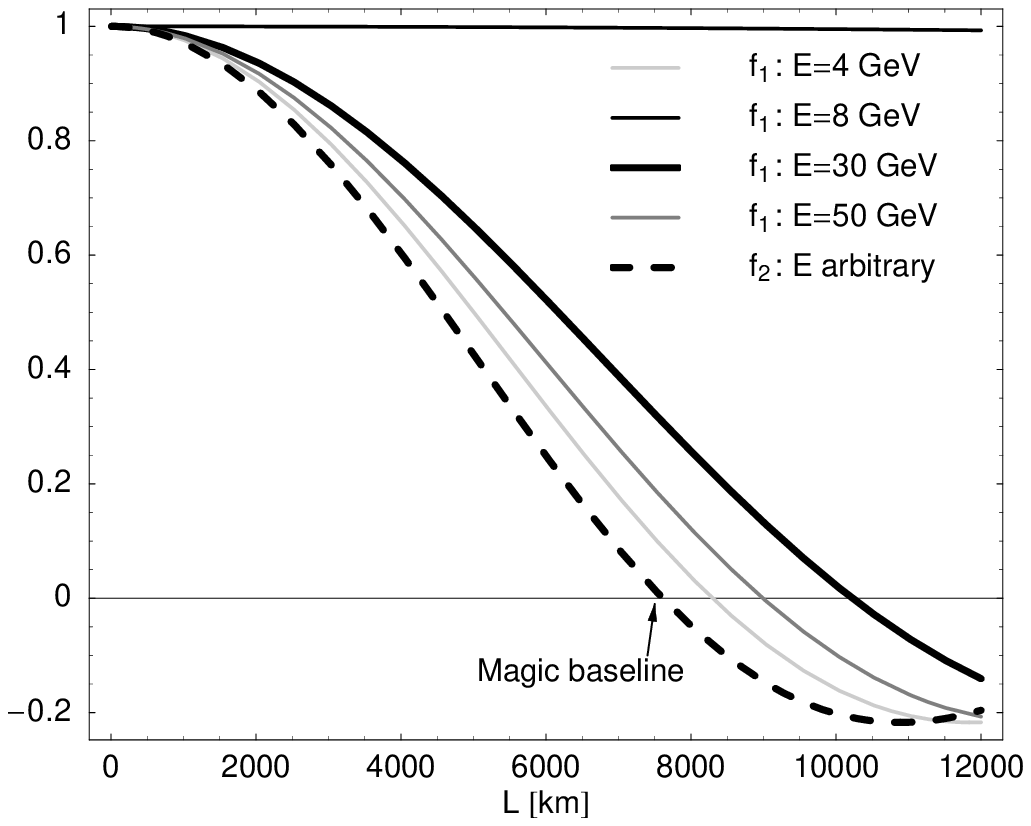}
\vspace*{-1.5cm}

\end{center}
\caption{\label{fig:terms} Factors $f_1$ (solid curves) and $f_2$ (dashed curve) from
\equ{PROBMATTER} as function of $L$ for different values of the energy. For this figure,
$\ldm = 0.0025 \, \mathrm{eV}^2$ and $\rho = 4.3 \, \mathrm{g}/\mathrm{cm^3}$ is used.}
\vspace*{-0.8cm}
\end{figure}
We show in \figu{terms} these two factors as function of $L$ for different values of $E$.
Obviously, $f_1$ depends on the energy and does not drop as function of $L$ close to the
matter resonance $\hat{A} \rightarrow 1$. However, $f_2$ does not depend on the
energy and has its first root at the ``magic baseline''.

This ``magic baseline''~\cite{Lipari:1999wy,
Huber:2003ak} is the first application from \equ{PROBMATTER}:
If we choose $f_2 \equiv 0$, the condition $\sin ( \hat{A} \Delta ) = 0$
evaluates to $\sqrt{2} G_F n_e(L) L = 2 \pi$ or $L_{\mathrm{magic}} \sim 7 \, 250 - 7 \, 500 \, \mathrm{km}$ for the first root. Obviously, the suppression of $f_2$ independent of
energy and oscillation parameters allows an almost correlation- and degenerate-free
measurement of $\stheta$ and the mass hierarchy,
whereas no information on $\deltacp$ can be obtained. As it has been demonstrated in
\Ref~\cite{Huber:2003ak} for a NF, the combination of the magic baseline with $L=3 \, 000 \, \mathrm{km}$ has excellent capabilities for the $\stheta$, mass hierarchy, and CP violation sensitivities down to $\stheta \simeq 10^{-4}$. In addition, the magic baseline can be used for a risk-minimized precision measurement of $\deltacp$~\cite{Huber:2004gg}.
\begin{figure}[t]
\begin{center}
\includegraphics[width=\columnwidth]{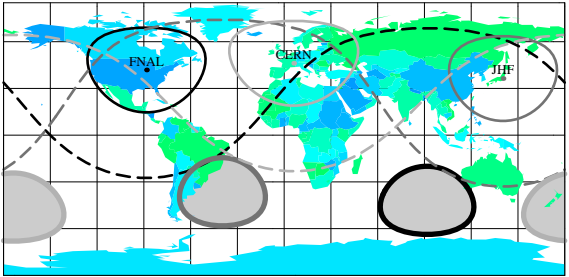}

\vspace*{-1cm}

\end{center}
\caption{\label{fig:basepl} Three of the major potential NF laboratories
and possible detector locations at $L=3 \, 000 \, \mathrm{km}$ (thin solid curves), $L=7 \, 250 \, \mathrm{km}$ (dashed curves),
and inner core crossing baselines (shaded regions) in the corresponding lab colors.}
\vspace*{-0.8cm}
\end{figure}
Potential magic baseline detector locations for three of the major potential NF laboratories can be found in \figu{basepl} on the dashed curves.

Another application with a different purpose can be read off from \equ{PROBMATTER} for $\stheta=0$: In this case, all but the last term vanish. It is an interesting feature that $f_2$, which dominates the magnitude of the remaining ``solar'' term, does not drop in vacuum ($f_2 \rightarrow 1$), but is very small in matter ($f_2 \rightarrow 0$ for $L \rightarrow \infty$).
Thus, the solar term is suppressed by the matter effect. One can use this effect for a
direct high confidence level verification of the MSW effect in Earth matter: For a $5 \sigma$ signal, a NF baseline $L > 6 \, 000 \, \mathrm{km}$ is required~\cite{Winter:2004mt}. As the most important observation, this result does
(compared to the mass hierarchy determination) not depend on $\stheta$ and even holds
for $\stheta=0$. Note that there are many potential detector locations for $L > 6 \, 000 \, \mathrm{km}$ -- in particular, the ``magic baseline'' satisfies this requirement (\cf, \figu{basepl}).

Finally, in the limit of large $\stheta$, \equ{PROBMATTER} reduces to the first term
as a first approximation. As one can read off from \figu{terms}, the in this case dominating
factor $f_1$ does not drop close to the matter resonance $\hat{A} \rightarrow 1$
even for very long baselines. Note that the $1/L^2$ drop of the flux is already factored out,
which means that the probability is proportional to $f_1^2$. However,
the further off the resonance, the stronger is the change in the probability. Therefore,
this factor becomes very sensitive to the matter density. In principle, it allows
a per cent level measurement of the absolute density of the Earth's core using a vertical
NF baseline, as it has been demonstrated in \Ref~\cite{Winter:2005ws} for $\stheta > 0.01$ including the
correlations with the neutrino oscillation parameters. In addition, as shown in \figu{basepl} for the baselines crossing the Earth's inner core, there are potential locations on land on the other sides of many of the major potential NF laboratories.

In summary, we have demonstrated that there are several potential applications of
``very'' long NF baselines, and additional ones, such as the mass hierarchy determination for $\stheta=0$, are under investigation~\cite{Theta0}. Therefore, we conclude that possible muon storage ring configurations should be studied which allow for the simultaneous or subsequent operation of such a baseline in combination with a shorter baseline. Furthermore, the decay tunnel slopes would be a major challenge for such an application.


\end{document}